\documentclass[a4paper]{article}

\usepackage{INTERSPEECH2020}

\usepackage{caption}
\usepackage[labelformat=simple]{subcaption}

\usepackage{multirow} 
\usepackage{booktabs} 
\usepackage{makecell}

\usepackage{CJKutf8} 

\usepackage{hyperref}
\usepackage{cite}

\title{The Sequence-to-Sequence Baseline for the Voice Conversion Challenge 2020: Cascading ASR and TTS}
\name{Wen-Chin Huang$^1$, Tomoki Hayashi$^1$, Shinji Watanabe$^2$, Tomoki Toda$^1$}
\address{$^1$Nagoya University, Japan\\$^2$Johns Hopkins University, USA}
\email{wen.chinhuang@g.sp.m.is.nagoya-u.ac.jp}

\begin{document}

\maketitle
\begin{abstract}
This paper presents the sequence-to-sequence (seq2seq) baseline system for the voice conversion challenge (VCC) 2020. We consider a naive approach for voice conversion (VC), which is to first transcribe the input speech with an automatic speech recognition (ASR) model, followed using the transcriptions to generate the voice of the target with a text-to-speech (TTS) model. We revisit this method under a sequence-to-sequence (seq2seq) framework by utilizing ESPnet, an open-source end-to-end speech processing toolkit, and the many well-configured pretrained models provided by the community. Official evaluation results show that our system comes out top among the participating systems in terms of conversion similarity, demonstrating the promising ability of seq2seq models to convert speaker identity. The implementation is made open-source at: \url{https://github.com/espnet/espnet/tree/master/egs/vcc20}.
\end{abstract}

\section{Introduction}

Voice conversion (VC) is a technique to transform the para-/non-linguistic characteristics included in a source speech waveform into a different one while preserving linguistic information \cite{VC, GMM-VC}. VC has great potential in the development of various new applications such as speaking aid devices for vocal impairments, expressive speech synthesis, silent speech interfaces, or accent conversion for computer-assisted language learning.

The aim of the voice conversion challenge (VCC)\footnote{\url{http://www.vc-challenge.org/}} is to better understand different VC techniques built on a freely-available common dataset to look at a common goal and to share views about unsolved problems and challenges faced by current VC techniques. The challenges focused on \textit{speaker conversion}, where VC models are built to automatically transform the voice identity. In the third version, VCC2020 \cite{vcc2020summary}, two new tasks are considered. The first task is \textit{semiparallel} VC within the same language, where only a small subset of the training set is parallel with the rest being nonparallel. The second task is \textit{cross-lingual} VC, where the training set of the source speaker is different from that uttered by the target speaker in language and content, thus nonparallel in nature. In conversion, the source speaker's voice in the source language is converted as if it was uttered by the target speaker while keeping linguistic contents unchanged.

%

It would be worth discussing two important factors when designing a VC system: data and model. First, from the data point of view, in either of the VCC2020 tasks, techniques for dealing with nonparallel data need to be developed. In the literature, a promising paradigm for nonparallel VC is through a recognition-synthesis framework. The idea is to first extract from the source speech the linguistic contents, followed by blending with the target speaker characteristics to generate the converted speech. Methods implementing this framework can be divided according to the type of linguistic representation. The first type encodes representations with an automatic speech recognition (ASR) model, where a popular choice is the phonetic posteriorgram (PPG) \cite{VC-PPG, VC-WNV-adapt}. A synthesis model is then trained to generate the voice of the target speaker. The second type usually employs an autoencoder-like model that estimates the recognizer and synthesizer simultaneously by implicitly factorizing the linguistic and speaker representations \cite{VAE-VC, VAE-GAN-VC, CHOU-NPVC, autovc, CDVAE-CLS-GAN}.

From the model point of view, we have witnessed how seq2seq models \cite{S2S} change the game in many research fields in only half a decade, and speech processing is no exception. Its application in VC is especially attractive since that compared to conventional frame-based methods that perform conversion frame-by-frame, seq2seq models can implicitly learn the complex alignment and relationship between the source and target sequences to generate outputs of various lengths. It is therefore a natural choice to convert prosody including the speaking rate and F0 contour, which is closely related to speaker characteristics. As a result, seq2seq based VC has been a promising approach in terms of conversion similarity \cite{ATT-S2S-VC, S2S-iFLYTEK-VC, S2S-NP-VC, VTN}.

\begin{figure}[t]
	\centering
	\includegraphics[width=0.45\textwidth]{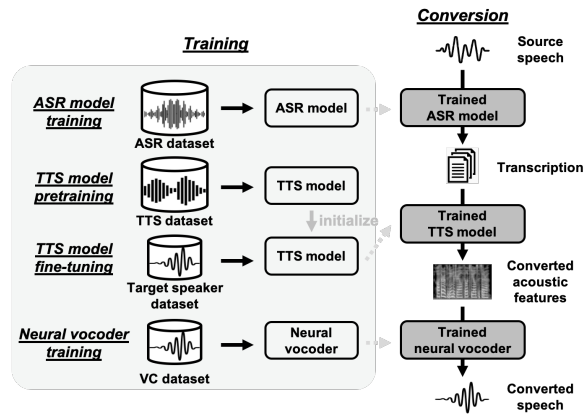} 
	\caption{The training and conversion processes of the \textsc{ASR+TTS} method. \label{fig:asr+tts}}
\end{figure}

In this paper, we describe the seq2seq baseline system for the VCC2020. Our system is a cascade of seq2seq-based ASR and TTS models, which we will refer to as \textsc{ASR+TTS}. A suitable baseline system should meet the following requirements:
\begin{itemize}
	\item The system should be a simple and easy-to-use starting ground for newcomers to base their work on.
	\item The system should be an open-source project made publicly available to benefit potential future researchers.
	\item The system should serve as a competitive benchmark.
\end{itemize}
With these goals in mind, we implemented the system using ESPnet, a well-developed open-source end-to-end (E2E) speech processing toolkit \cite{espnet, espnet-tts}, and made as much use of publicly available datasets as possible. Although it is generally believed that simply cascading systems to perform a certain task is inferior to an end-to-end model, benefitting from recent advances in ASR and TTS, as well as efforts such as implementation and hyperparameter tuning which are dedicated by the open-source community, we will show that our system is not only easy to use but serves as a strong competing system in the VCC2020.

\section{System Overview}

A naive approach for VC is a cascade of an ASR model and a TTS model.
Although this method is not new, by revisiting this method using seq2seq models, we can model the prosody such as pitch, duration, and speaking rate, which is usually not well considered in the literature.
Conceptually speaking, the ASR model acts like a speaker normalizer that first normalizes the input speech such that attributes of the source speaker are filtered out and only the linguistic content remains. 
Then, the TTS model functions to add speaker information to the recognition result so that the converted speech sounds like the target speaker.

Our system, as depicted in Figure~\ref{fig:asr+tts}, consists of three modules: a speaker-independent ASR model, a separate speaker-dependent TTS model for each target speaker, and a neural vocoder that synthesizes the final speech waveform.

\noindent{\textbf{ASR model.}} ASR models are usually trained with a multi-speaker dataset, thus speaker-independent in nature. For both tasks 1 and 2, the source speech is always English, so an English transcription is first obtained using the ASR model. 

\noindent{\textbf{TTS model.}} In the TTS literature, it is a common practice to train in a speaker-dependent manner rather than training speaker-independently since the former usually outperform the latter. However, the size of the training set of each target speaker in VCC2020 is too limited for seq2seq TTS learning. In light of this, we employ a pretraining-finetuning scheme that first pretrains on large TTS datasets followed by fine-tuning on the limited target speaker dataset\cite{semi-speech-synthesis} . This allows us to successfully train on even approximately 5 minutes of data.

\noindent{\textbf{Neural vocoder.}} In recent years, neural waveform generation modules (also known as vocoders) have brought significant improvement to VC. In this work, we use the Parallel WaveGAN (PWG) \cite{parallel-wavegan}, since it enables high-quality, real-time waveform generation. An open-source implementation\footnote{\url{https://github.com/kan-bayashi/ParallelWaveGAN}} is adopted and we integrated it with ESPnet.

Our implementation was built upon the E2E speech processing toolkit ESPnet \cite{espnet, espnet-tts}, which provides various useful utility functions and properly tuned pretrained models.

\section{ASR Implementation}

\subsection{Data}
Since the input is always English, we used the Librispeech dataset \cite{librispeech}, which contained 960 hours of English speech data from over 2000 speakers.

\subsection{Model}
The backbone of the ASR model was the Transformer \cite{transformer, transformer-asr, transformer-asr-ctc-lm}. The model was trained in an end-to-end fashion using a hybrid CTC/attention loss \cite{ctc-attention}, and a recurrent neural network based language model (RNNLM) was used for decoding. We directly used a pretrained model (including the RNNLM) provided by ESPnet.

\begin{table}[t]
	\centering
	\caption{The TTS training datasets in task 2. "phn" and "char" stand for phoneme and character, respectively.}
	
	\centering
	\begin{tabular}{ c c c c c}
		\toprule
		Lang. & Dataset & Spkrs & Hours & Input \\
		\midrule
		Eng. & M-AILABS \cite{M-AILABS} & 2 & 32 & phn or char \\
		Ger. & M-AILABS \cite{M-AILABS} & 5 & 190 & char \\
		Fin. & CSS10 \cite{css10} & 1 & 10 & char \\
		Man. & CSMSC \cite{csmsc} & 1 & 12 & pinyin \\
		\bottomrule
	\end{tabular}
	\label{tab:task2-tts-datasets}
\end{table}

\begin{figure}[t]
	\centering
	\includegraphics[width=0.48\textwidth]{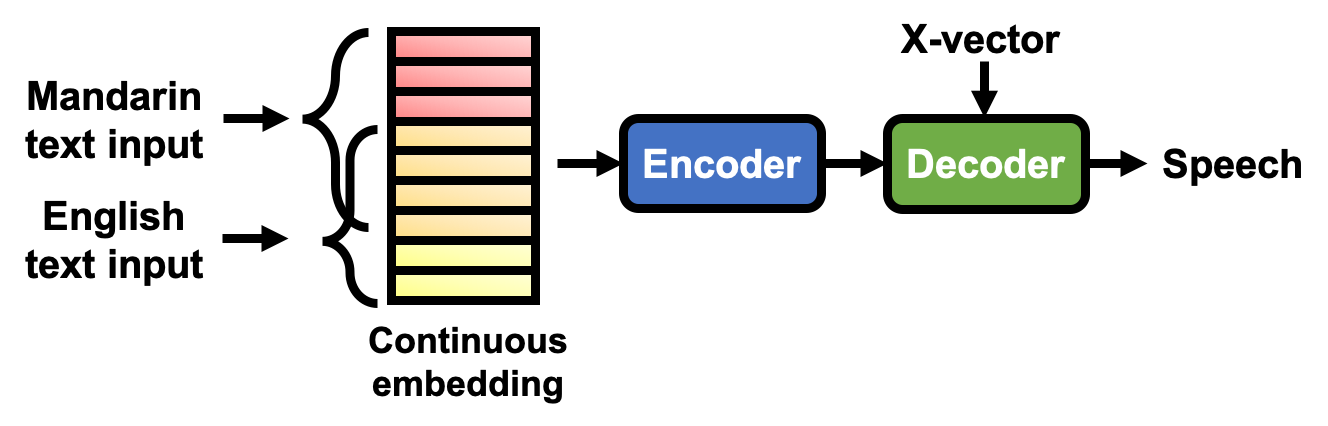} 
	\caption{Illustration of the bilingual TTS used in task 2. \label{fig:bilingual-tts}}
\end{figure}

\section{TTS Implementation}

We are faced with a harder challenge in implementing the TTS model. In task 2, the input language is different from the languages of the training data. In other words, the TTS model needs to lean the voice of an unseen language. This is sometimes referred to as cross-lingual voice cloning \cite{cross-lingual-voice-cloning, mix-lingual-tts-with-monolingual}. As there has not been a standard, promising protocol especially when only five minutes of training data is available, we adopt a simple method that constructs x-vector \cite{x-vector} based, bilingual TTS models by pretraining with corpora of English and the target language and finetuning with the target language.

\subsection{Data}
\label{ssec:task2-data}

The target language for task 1 is English, so for pretraining, we used the multi-speaker LibriTTS \cite{libritts} dataset, which contained around 250 hours of English data from over 2000 speakers. In task 2, the target languages are German, Finnish, and Mandarin. Considering the open-source ability, we wish to avoid using commercial or private datasets. Unfortunately, under such constraint, there is not much choice, and the available datasets at the time we developed the system were large but contained only data from a single speaker or a few speakers, as shown in Table~\ref{tab:task2-tts-datasets}. Although it has been shown that combining imbalanced multi-speaker datasets improves performance \cite{multispeaker-tts-imbalanced-data}, this effect remains unknown in the cross-lingual setting. To this end, for the English data, we decided to use not the LibriTTS dataset which has many speakers yet a small amount of data per speaker, but the M-AILABS dataset \cite{M-AILABS}, which has a large amount of data from a few speakers only. Finally, since the task 2 datasets were of different sampling rates, we doswnsampled all task 2 data to 16 kHz. As for the x-vector extractor, the Kaldi toolkit was used and the model was pretrained on VoxCeleb \cite{voxceleb}.

\begin{figure*}[t]
    \centering
	
	\begin{subfigure}{0.44\textwidth}
		\centering
  		\includegraphics[width=\textwidth]{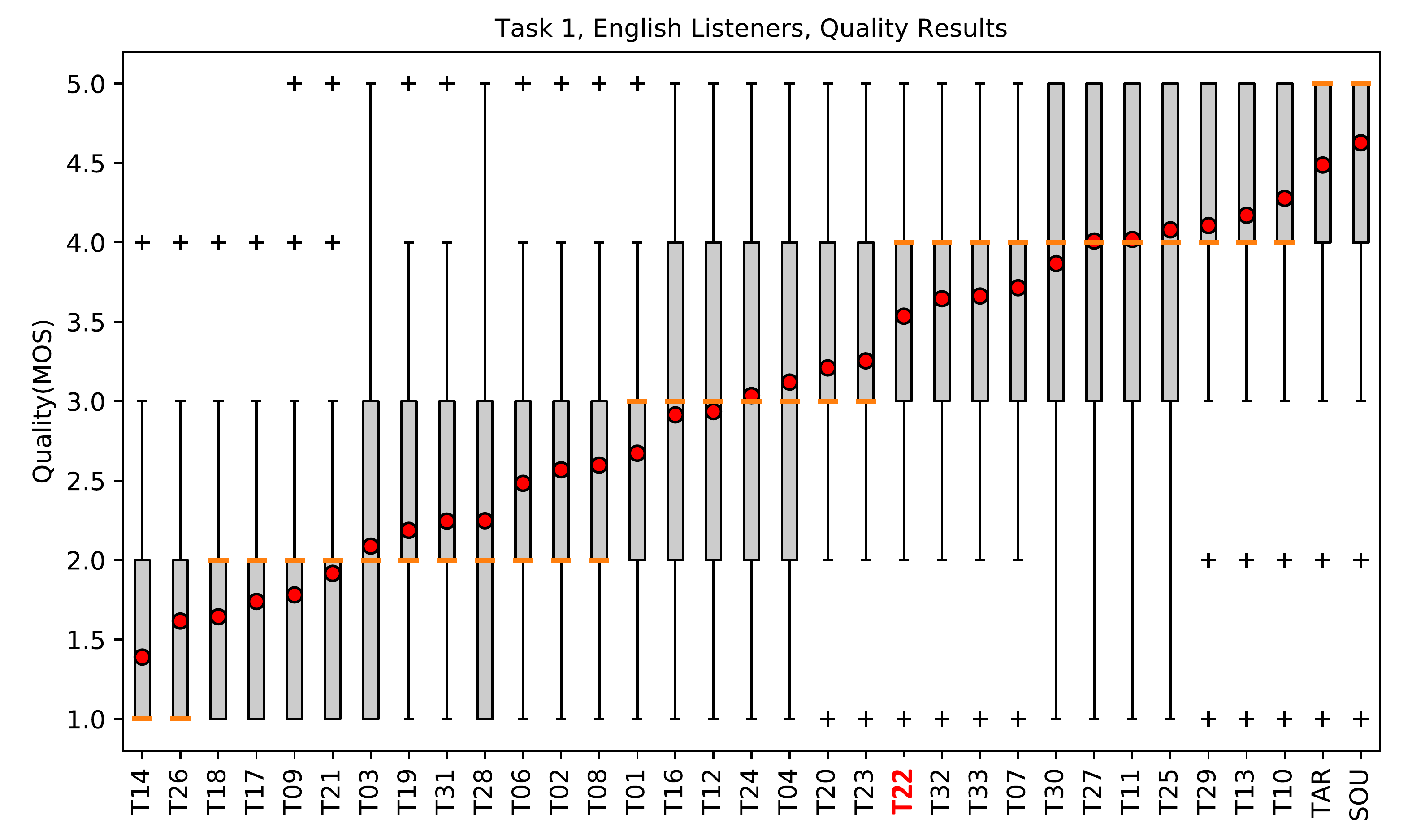}
		\caption{Naturalness results for task 1.}
   		\label{fig:nat-task1}
	\end{subfigure} %
	\begin{subfigure}{0.44\textwidth}
		\centering
  		\includegraphics[width=\textwidth]{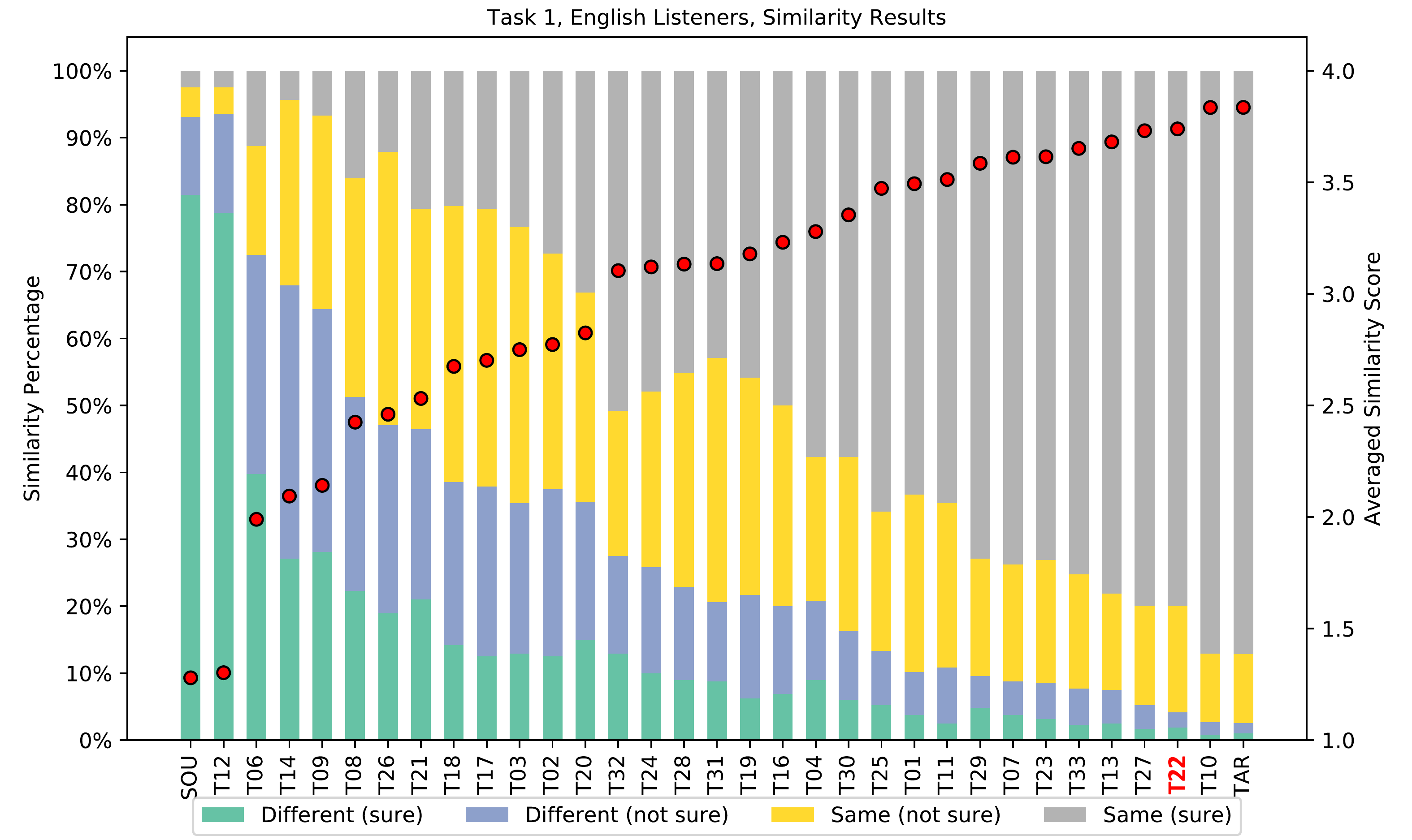}
		\caption{Similarity results for task 1.}
   		\label{fig:sim-task1}
	\end{subfigure}
	
	\vspace{0.5cm}
	
	\begin{subfigure}{0.44\textwidth}
		\centering
  		\includegraphics[width=\textwidth]{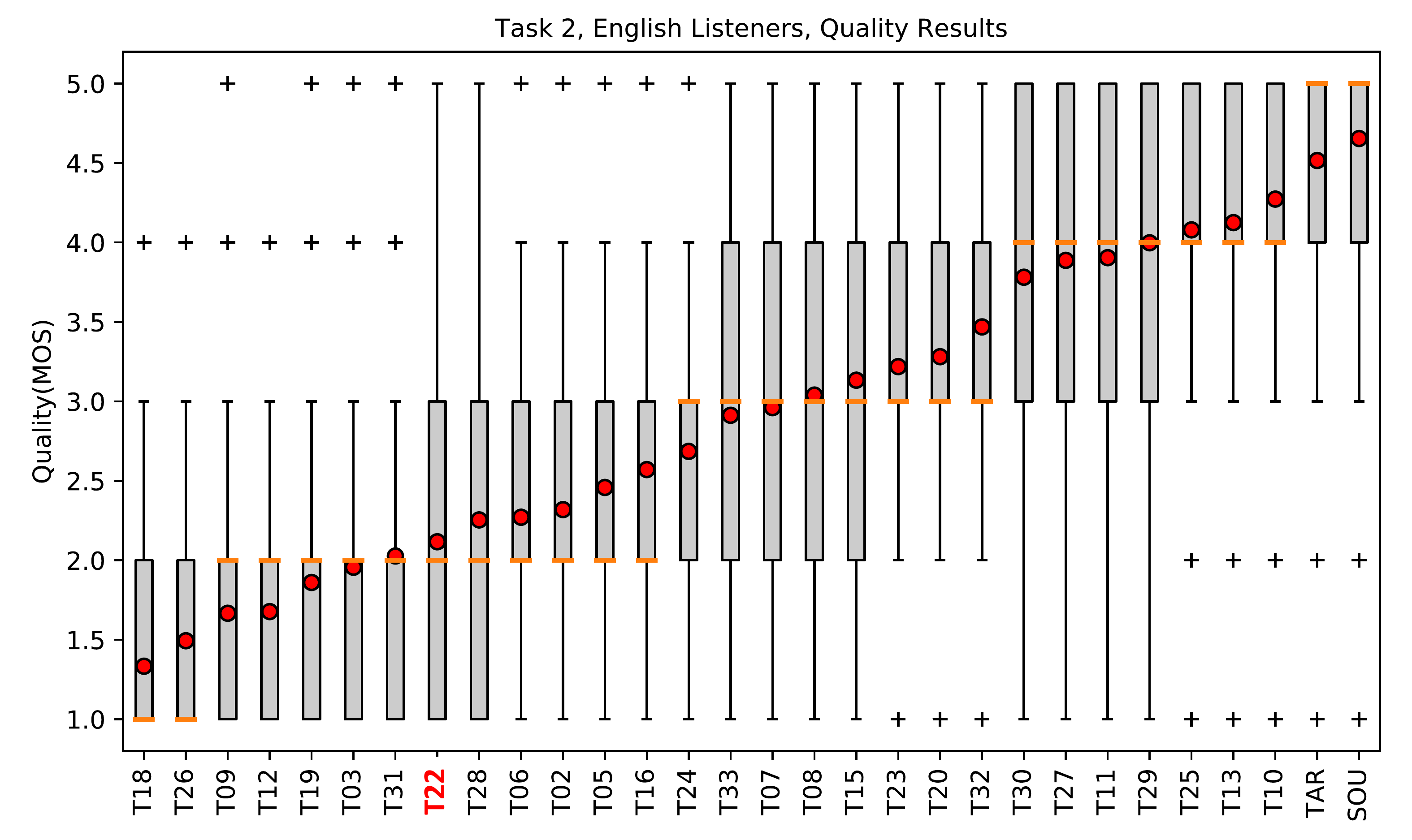}
		\caption{Naturalness results for task 2.}
   		\label{fig:nat-task2}
	\end{subfigure} %
	\begin{subfigure}{0.44\textwidth}
		\centering
  		\includegraphics[width=\textwidth]{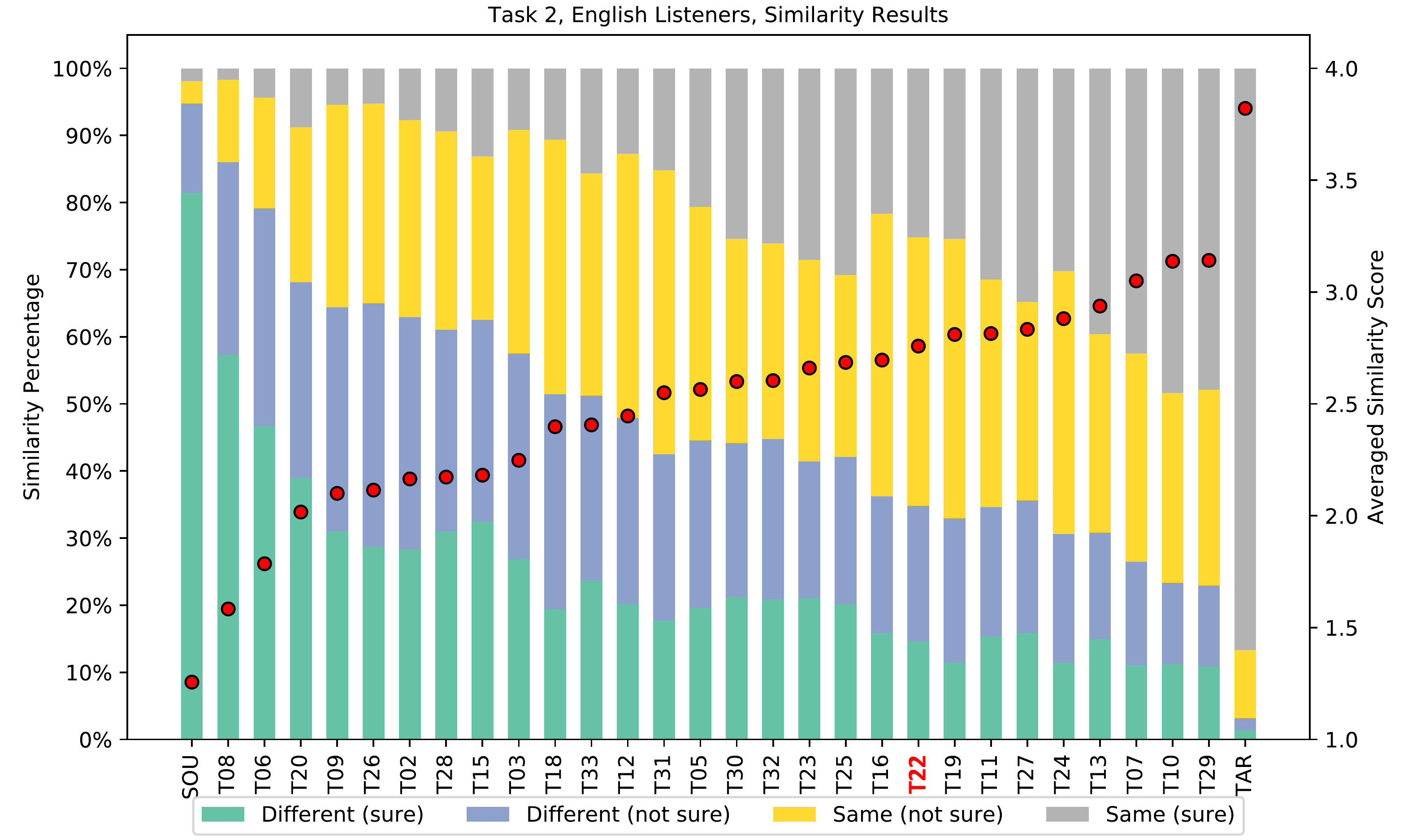}
		\caption{Similarity results for task 2.}
   		\label{fig:sim-task2}
	\end{subfigure}
	
	\caption{Official evaluation results of the VCC2020. Our system is T22, as emphasized in red.}
	\label{fig:results}
\end{figure*}

\subsection{Model}
\label{ssec:task2-model}

We used an x-vector \cite{x-vector} based multi-speaker TTS model \cite{asv-tts} with a Transformer backbone \cite{transformer-tts}. The input was a linguistic representation sequence, and the output was the mel filterbank sequence extracted from the (optionally downsampled) waveform. In task 1, since the input is always English, the model simply takes English characters an input.

However, in task 2, it is nontrivial to decide the input representation since it is often language-dependent. For example, there is no overlap in the text representation between Mandarin and English \cite{cross-lingual-voice-cloning}. When we finetune a pretrained model for a Mandarin speaker, since the Mandarin corpus does not contain English words, the model has no clue how the target speaker pronounce English words. This mismatch may cause quality degradation. Below, we describe how we alleviate this issue.

We used a shared input embedding space when training the bilingual TTS model. In neural TTS, the input embedding look-up table is a projection from discrete input symbols to continuous representation and is trained with the rest of the model by backpropagation. It is useful in that the model can implicitly learn how to pronounce each input token, such that different tokens with a similar pronunciation can have a similar embedding. The assumption here is that there is an overlap between the input representations of the two languages. For example, if we train a Mandarin/English TTS model, the "ah" phoneme in English and "a" pinyin representation may have similar embeddings. As a result, even if only "a" is seen during training, by learning how the target speaker pronounces such vowel, the model may still know how to pronounce "ah".

For the Mandarin/English TTS, we used phonemes and pinyin as input, while for the Finnish/English and German/English TTS, we used characters as input. In the finetuning stage, the parameters are updated using the training utterances of the target speaker, except that the embedding lookup table in Figure~\ref{fig:bilingual-tts} is fixed. 

\section{Neural Vocoder Implementation}
\label{sec:vocoder}

The PWG had a non-autoregressive (non-AR) WaveNet-like architecture and was trained by jointly optimizing a multi-resolution spectrogram loss and a waveform adversarial loss \cite{parallel-wavegan}. The input was mel filterbank and the output was raw waveform. For each task, we trained a separate PWG using the training data from all available speakers. In other words, data of 8 and 10 speakers were used to train PWGs for tasks 1 and 2, respectively. Notably, in task 2, although the mel filterbanks were extracted from 16kHz waveform as mentioned in Sections~\ref{ssec:task2-data} and~\ref{ssec:task2-model}, we still map them to 24kHz waveform in training, as the quality degradation from such mismatch has shown to be acceptable \cite{libritts}.


\section{Challenge Results}

\subsection{VCC2020 Dataset}

The VCC2020 database had two male and two female English speakers as the source speakers. For task 1, two male and two female English speakers were chosen as the target speakers, and one male and one female for each of Finnish, German, and Mandarin in task 2. Each of the source and target speakers has a training set of 70 sentences, which is around 5 minutes of speech data. Note that in task 1, the target and source speakers have 20 parallel sentences, where the rest 50 sentences are different. The test sentences for evaluation are shared for tasks 1 and 2 with a number of 25.

\subsection{Evaluation protocol}

The VCC2020 organizing committee conducted a large-scale subjective test on all submitted systems for both tasks 1 and 2. The evaluations included naturalness and similarity tests. In the naturalness test, a five-point mean opinion score (MOS) test was adopted, where listeners were asked to rate the naturalness of each speech clip from 1 to 5. In the similarity test, listeners were presented with a converted and a ground truth target utterance, and they were asked to decide whether or not the two utterances were spoken by the same person on a four-point scale. Figure~\ref{fig:results} shows the overall results\footnote{Although the official report contained results from Japanese and English listeners, here we only report results of English listeners since the two listener groups share a similar tendency.}.

\subsection{Task 1 Results}

Figures~\ref{fig:nat-task1} and~\ref{fig:sim-task1} show the overall results for task 1. For naturalness, our system received a MOS score of about 3.5, which ranks 11 out of all the 31 submitted systems in task 1.  This shows that, as many systems are specifically designed for VC, simply combining state-of-the-art ASR and TTS systems can already achieve competitive results, thanks to the well-developed technologies in the two research fields. The performance gap between our system and the superior teams may come from the difficulty of finetuning the TTS model with only 70 utterances. As for similarity, our system had a similarity score around 90\%, which means that about 90\% of the converted utterances were considered spoken by the same target speaker by the participants. This made our system rank second among all teams, which serves as strong evidence of the superiority of seq2seq models when it comes to converting speaker identity.

\subsection{Task 2 Results}
\label{ssec:task2-results}

Figures~\ref{fig:nat-task2} and~\ref{fig:sim-task2} show the overall results for task 2. For naturalness, our system had a MOS score of about 2.0, ranking 21 out of all the 28 submitted systems in task 2, which is a lot worse than the performance in task 1. On the other hand, our system ranked 9 among the 28 teams in the similarity test. Looking at these two results, it can be inferred that our system can still well capture the speaker characteristics thanks to the power of seq2seq modeling, but suffer from a severe quality degradation. This is possibly owing to the limited training data and the lack of pretraining data, as well as the difficulty of handling the cross-lingual data using the overly-simple TTS model we implemented.

\begin{table}[t]
	\centering
	\caption{Character/word error rates (CER/WER) (\%) calculated using a pretrained ASR model. The scores are averaged over all target speakers.
	}
	
	\centering
	\begin{tabular}{ c c c c c c c}
		\toprule
		& \multicolumn{2}{c}{Input} & \multicolumn{2}{c}{Task 1} & \multicolumn{2}{c}{Task 2} \\
		\cmidrule(lr){2-3} \cmidrule(lr){4-5} \cmidrule(lr){6-7}
		Source & CER & WER & CER & WER & CER & WER \\
		\midrule
		SEF1 & 2.9 & 6.5 & 12.1 & 22.1 & 19.9 & 34.3\\
		SEF2 & 1.4 & 3.7 & 12.6 & 22.7 & 21.4 & 36.2\\
		SEM1 & 0.2 & 0.9 & 14.2 & 20.1 & 20.3 & 36.8 \\
		SEM2 & 2.9 & 7.5 & 18.5 & 30.9 & 22.7 & 38.0 \\
		\bottomrule
	\end{tabular}
	\label{tab:asr-results}
\end{table}

\section{Analysis on Linguistic Contents}

A potential threat of the cascading paradigm is that error in early stages might propagate to downstream models. In our proposed method, the recognition failure in the first ASR stage might harm the linguistic consistency in VC. We examine this phenomena by measuring the intelligibility with an off-the-shelf Transformer-ASR model trained on LibriSpeech, which is provided in ESPnet.

Table~\ref{tab:asr-results} shows the ASR results. First, the error rates on the input source speech were not severe as they are similar to that on the test set of LibriSpeech. However, the scores of the converted speech are much worse, indicating that the imperfect TTS modeling is the main cause of intelligibility degradation. We also observe that the error rates of task 2 are much higher than that of task 1, which is consistent with the results in Section~\ref{ssec:task2-results}.

\section{Conclusion and Discussion}

This paper described the seq2seq baseline system of the VCC2020, including the intuition, system design, training datasets, and results. Built upon the E2E, seq2seq framework, our \textsc{ASR+TTS} baseline served as a simple starting point and a benchmark for participants. Subjective evaluation results released by the organizing committee showed that our system is a strong baseline in terms of conversion similarity, confirming the effectiveness of seq2seq modeling. The results also demonstrate the naive yet promising power of combining state-of-the-art ASR and TTS models. Yet, there is still much room for improvement, and below we discuss several possible directions that might be addressed in an advanced version.

\noindent{\textbf{Enhance the pretraining data.}}
As stated in Section~\ref{ssec:task2-data}, there was not much choice for pretraining data in task 2 under the open-source constraint. Using a multi-speaker pretraining dataset as in task 1 might improve the performance. Also, using datasets with a higher sampling rate can also improve the quality of the vocoder.

\noindent{\textbf{Utilize linguistic knowledge.}}
One principal of E2E learning to use as less domain-specific knowledge as possible, That is to say, the system performance is expected to be improved when such knowledge is utilized. For example, as reported in \cite{cross-lingual-voice-cloning}, using phoneme inputs can greatly improve multi-lingual TTS systems, but we could not do so in task 2 due to the unfamiliarity with target languages such as Finnish and German.

\noindent{\textbf{Select an advanced multi-speaker TTS model.}}
The multi-speaker TTS model \cite{asv-tts} we adopted was a rather naive one, and a more state-of-the-art model like \cite{GMVAE} might improve the performance.

\noindent{\textbf{Improve the neural vocoder.}}
We adopted a non-AR neural vocoder for fast generation, but it is generally believed that AR ones are still superior. As this is a popular research field, it is expected that real-time neural vocoders maintaining the output quality will soon be developed. Also, finetuning the vocoders can further improve the performance, as stated in Section~\ref{sec:vocoder}.

\section{Acknowledgement}

This work was supported in part by JST, CREST Grant Number JPMJCR19A3 and JSPS KAKENHI Grant Number 17H06101.

\bibliographystyle{IEEEtran}
\bibliography{ref}

\end{document}